\documentclass[a4paper]{JHEP3}
\usepackage{graphicx}

\newcommand{\bm}[1]{\mbox{{\boldmath $#1$}}}
\title{A criterion for lattice supersymmetry: cyclic Leibniz rule}
\author{Mitsuhiro Kato\\
        Institute of Physics, University of Tokyo, Komaba, 
        Meguro-ku, Tokyo 153-8902, Japan\\
        E-mail: \email{kato@hep1.c.u-tokyo.ac.jp}
        }
\author{Makoto Sakamoto\\
        Department of Physics, Kobe University, Nada-ku, 
        Hyogo  657-8501, Japan\\
        E-mail: \email{dragon@kobe-u.ac.jp}
        }
\author{Hiroto So\\
        Department of Physics,  Ehime University, Bunkyou-chou 2-5, 
        Matsuyama 790-8577, Japan\\
        E-mail: \email{so@phys.sci.ehime-u.ac.jp}
        }
\abstract{
It is old folklore that the violation of Leibniz rule on a lattice is
an obstruction for constructing a lattice supersymmetric model. 
While it is still true for full supersymmetry, we
show that a slightly modified form of the Leibniz rule, which we call
cyclic  Leibniz rule (CLR), is actually a criterion for the
existence of partial lattice supersymmetry. 
In one dimension, we find sets of lattice
difference operator and field multiplication smeared over lattice
 which satisfy the CLR under
some natural assumptions such as translational invariance and
locality. Thereby we construct a model of supersymmetric lattice
quantum mechanics without spoiling locality. 
The CLR relation is coincident with the condition
that the vanishing of the so-called surface term in the construction
by lattice Nicolai map. We can construct superfield formalism
 with arbitrary superpotential.
This also enables us to apply safely a localization
technique to our model, because the kinetic term and the interaction
terms of our model are independently invariant under the supersymmetry
transformation. A preliminary attempt in finding a solution for the higher
dimensional case is also discussed.

}
\keywords{lattice, Leibniz rule, supersymmetry}
\preprint{\hspace{-1mm}UT-Komaba/13-2, KOBE-TH-13-03, EHIME-TH-8}
\begin{document}
%
%
%

%
\section{Introduction}
%
%
%

Supersymmetry (SUSY) is not only a candidate for physics beyond the standard model 
but also a possible framework (or a part of it) with definite ultraviolet behaviors in quantum field theories.  
If we can construct supersymmetric theories nonperturbatively, 
we shall obtain deeper understandings of vacuum property and 
 other  nonperturbative behaviors\cite{Feo2003,Kaplan2004,Kaplan-Catterall}. 
Although the SUSY algebra is generally incompatible with a discrete lattice space,  
there is no obstruction to formulate a free SUSY theory on lattice, 
if we replace an infinitesimal translation with a lattice translation.  

Our real difficulty  to overcome 
is Leibniz rule (LR) on lattice  in  interacting SUSY theories. 
Unfortunately, the no-go theorem for the rule on lattice has been proved in our previous papers\cite{KSS,KSS2}. 
The essential point of the proof is the incompatibility of the translational invariance,
 the locality and the LR on lattice. Under a certain circumstances, SLAC type difference operator  satisfies the LR on lattice\cite{KSS,SLAC} but it is non-local. 

As another approach, we could utilize a Nicolai map for constructing 
a lattice SUSY theory\cite{Sakai-Sakamoto}.
For supersymmetric quantum mechanics, a lattice model with 
a Nicolai map was proposed in Ref.\cite{LSUSY1},
%
\begin{equation}
S_{N} = \sum_{n}\frac{1}{2}
        \Bigl( (\Delta\phi)_{n} - W_{n}(\phi) \Bigr)^{2}
        + \sum_{m,n} i\bar{\psi}_{m}
          \left( \Delta_{mn} + \frac{\partial W_{n}(\phi)}{\partial \phi_{m}}
          \right) \psi_{n}\,,
\label{LSUSY_action1}
\end{equation}
%
where $\Delta_{mn}$ is a difference operator and $W_{n}(\phi)$
is an arbitrary function of $\phi_{m}$.
The lattice action (\ref{LSUSY_action1}) possesses
an exact supersymmetry\cite{LSUSY2,LSUSY3} but 
contains the unconventional term
%
\begin{equation}
- \sum_{n}(\Delta\phi)_{n} W_{n}(\phi)\,.
\label{surface_term}
\end{equation}
%
It should be noticed that the above term is absent from the
action of supersymmetric quantum mechanics and  
would vanish in the continuum as a surface term.
It cannot be, however, written into the form of total divergences
due to the lack of the Leibniz rule on lattice.
The author\cite{LSUSY4} has discussed,
in a topological point of view, an off-shell action
%
\begin{equation}
S_{B} = \sum_{n}\frac{1}{2}(B_{n})^{2}
       + \sum_{n}iB_{n}\Bigl( (\Delta\phi)_{n} - W_{n}(\phi) \Bigr)
        + \sum_{m,n} i\bar{\psi}_{m}
          \left( \Delta_{mn} + \frac{\partial W_{n}(\phi)}{\partial \phi_{m}}
          \right) \psi_{n}\,,
\label{LSUSY_action2}
\end{equation}
%
which reduces to (\ref{LSUSY_action1})
after eliminating the auxiliary field $B_{n}$.
The above lattice action was rederived in a superfield 
formulation\cite{LSUSY5}.
Although the lattice action has an exact fermionic symmetry,
it may be regarded as a lattice version of the stochastic
action\cite{stochastic} rather than the 
standard supersymmetric quantum mechanics,
%
\begin{equation}
S_{\rm{SQM}} 
  = \int dt 
     \left\{ \frac{1}{2}\left(\frac{d\phi}{dt}\right)^{2}
             + \frac{1}{2} F^{2} + iF W(\phi)
             + i\bar{\psi} \left( \frac{d}{dt} 
             + \frac{\partial W(\phi)}{\partial \phi}\right) \psi \right\}
\label{SQM_action}
\end{equation}
%
with a Euclidean time $t$.
Hence, it will be worth looking for a lattice formulation
which leads to a lattice version of 
$S_{\rm{SQM}}$ and also does not rely on the existence of 
local Nicolai maps without unwanted terms like (\ref{surface_term}).

In this paper, we propose a novel criterion for the lattice SUSY realization which we call cyclic Leibniz rule (CLR), that is not a genuine LR on lattice but still
a natural lattice analog of the Leibniz rule in the continuum theories in the sense that it is reduced to the corresponding relation in the continuum limit. We construct sets of the lattice difference operator and the associated symmetric field product which satisfy the CLR. With these set of difference operator and field product, we formulate the concrete example of lattice SUSY quantum mechanics. Although only a part of SUSY could 
be realized exactly, yet they have several advantages over the existing approaches.
One of the advantages is that our lattice model is free from
the \lq\lq surface\rq\rq term (\ref{surface_term}) and is just
a lattice analog of the standard supersymmetric action (\ref{SQM_action}).
This property with CLR allows us to construct a superfield formulation,
whose realization is different from that given in Ref.\cite{LSUSY5}.
In the action of our model the kinetic term and the interaction terms are separately invariant under SUSY transformation. Because of this property we can apply a localization technique to obtain some exact results in our model.
Since our lattice formulation mimics the continuum supersymmetric quantum
mechanics and does not rely on the existence of local Nicolai maps,
it may shed a new light on the construction of higher-dimensional
supersymmetric lattice models.

In section 2, we propose the cyclic Leibniz rule (CLR) as a criterion for the SUSY by using a concrete example, for definiteness, of $D=1$ ${\cal N}=2$ Wess-Zumino model on lattice.   
The lattice difference opertors and associated symmetric product is found as a solution to the CLR.  
Based on these, in section 3, we construct exactly (half-)supersymmetric quantum mechanics on lattice with arbitrary interactions.
The Witten index is exactly calculated by the Nicolai map and the localization technique.
Multicomponent cases are discussed in section 4 and we find a local solution  which is a natural generalization of one-component case.  
In section 5, we give an attempt to a higher-dimensional case letting spinor index aside, while we so far could not find any suitable solution of the CLR for euclidean field theories.
Section 6 is devoted to conclusion  and discussions.
In appendies, we find the general solution of CLR with locality, the properties of the associated product,  
and  explicit multi-field product examples based on CLR.  

\vspace{5mm}
%
%
%
\section{A criterion for lattice supersymmetry}
%
%
%

In this section, we begin by recapitulating the difficulty in constructing interacting supersymmetric theories on lattice where the important key concept is the Leibniz rule.
Our proposal to avoid the difficulty will appear subsequently in this section. 

For concreteness, we consider $D=1$ model, {\it i.e.}~supersymmetric quantum mechanics, which consists of a multiplet
\begin{equation}
\Big(\phi_n,\psi_n, F_n\Big)
\label{multiplet-0}
\end{equation}
\noindent
where $\phi$ is a real scalar, $F$ is a real auxiliary field and $\psi$ is a complex fermion.  Index $n$ stands for a lattice site. $na$ is a coordinate 
but hereafter we set lattice constant $a=1$ for simplicity. 
The ${\cal N}=2$ SUSY transformation is described as 
\begin{eqnarray}
\delta \phi_n &=& \epsilon\bar{\psi}_n-\bar{\epsilon}\psi_n,  \nonumber\\
\delta \psi_n &= &\epsilon (i (\Delta\phi)_n + F_n),  \nonumber\\
\delta \bar{\psi}_n &=& \bar{\epsilon} (-i (\Delta\phi)_n + F_n),  \nonumber \\
\delta F_n &=& -\epsilon i (\Delta\bar{\psi})_n-\bar{\epsilon}i(\Delta\psi)_n, 
\label{SUSY-1}
\end{eqnarray}
\noindent
where  a difference operator $\Delta$ is defined as 
\begin{equation}
(\Delta A)_n \equiv \sum_m  \Delta_{nm} A_m, ~~\sum_m \Delta_{nm}=0,
\label{diff}
\end{equation}
\noindent
for arbitray lattice field $A_n$. The second relation states that the difference of a constant field vanishes.
The algebra  follows from the above definition:
\begin{equation}
[\delta_1,\delta_2]\Big(\phi_n, \psi_n, F_n \Big)  = 2i(\bar{\epsilon}_1\epsilon_2-\bar{\epsilon}_2\epsilon_1)
\Big( (\Delta\phi)_n,  (\Delta\psi)_n, (\Delta F)_n\Big).
\label{algebra-1}
\end{equation}
\noindent

The kinetic terms $S_0$ and the mass term $S_m$ of the free action
\begin{equation}
S_0= \sum_n \big(   \frac{1}{2}(\Delta\phi)_n^2  + i\bar{\psi}_n(\Delta\psi)_n+\frac{1}{2}F_n^2 \big), 
\label{free-1}
\end{equation}
\noindent
\begin{equation}
S_m=im\sum_n \big( F_n\phi_n+\bar{\psi}_n\psi_n \big) 
\label{mass-1}
\end{equation}
\noindent
are  invariant under (\ref{SUSY-1}) provided
\begin{equation}
\Delta_{nm}=-\Delta_{mn}
\label{symmetric diff}
\end{equation}
\noindent
which  is a generalized symmetric difference operator.

The obstruction for SUSY realization on lattice appears 
in the interaction parts. For an illustration we consider here cubic interaction such as  
\begin{eqnarray}
S_{{\rm int}}= i\frac{g}{2}\sum_{\ell m n} \big(  M_{\ell mn}F_{\ell}\phi_m\phi_n + 2N_{\ell mn}\phi_{\ell}\bar{\psi}_m\psi_n \big) ,
\label{interaction-1}
\end{eqnarray}
\noindent
where $g$ is a coupling constant, and $M_{\ell mn}$ and $N_{\ell mn}$  are  coefficients for  fields products. We can impose $M_{\ell mn} = M_{\ell nm}$ without the loss of generality.  
Under (\ref{SUSY-1}), the interaction terms are transformed  as follows, 
\begin{eqnarray}
\delta S_{{\rm int}} &=& i\frac{g}{2}\sum_{\ell m n} \Big(  
-2\bar{\epsilon}N_{\ell mn}\psi_{\ell} \bar{\psi}_m \psi_n
 +2 \epsilon N_{\ell mn}\bar{\psi}_{\ell}\bar{\psi}_m\psi_n
\nonumber \\
&&
-2\bar{\epsilon}( M_{\ell mn}-N_{m \ell n})F_{\ell} \psi_m \phi_n
 +2 \epsilon(M_{\ell mn}-N_{n m \ell})F_{\ell}\bar{\psi}_m\phi_n
\nonumber \\
&&
-i\bar{\epsilon}(M_{\ell mn} (\Delta\psi)_{\ell} \phi_m  \phi_n 
+  2N_{\ell mn} \phi_{\ell}  (\Delta\phi)_m \psi_n)  \nonumber \\
                     & & -i \epsilon (M_{\ell mn}(\Delta\bar{\psi})_{\ell} \phi_m \phi_n 
                     +2N_{\ell mn} \phi_{\ell} \bar{\psi}_m  (\Delta\phi)_n) 
                     \Big )  .
\label{tranform of int-1}
\end{eqnarray}
\noindent
In order to accomplish the invariance under  (\ref{SUSY-1}), we must require 
\begin{eqnarray}
N_{\ell mn} &=& N_{m \ell n}, \nonumber  \\
M_{\ell mn}&=&N_{nm \ell}, \nonumber \\
\sum_{k}(\Delta_{k\ell}M_{kmn} &+& N_{n\ell k}\Delta_{km} + N_{m\ell k}\Delta_{kn})=0 . 
\label{condition-1}
\end{eqnarray}
for $\epsilon$ invariance and
\begin{eqnarray}
N_{\ell mn}&=&N_{n m \ell},   \nonumber \\
M_{\ell mn} &=& N_{m \ell  n},  \nonumber \\
\sum_{k}(\Delta_{k \ell}M_{kmn} &+& N_{nk\ell}\Delta_{km} + N_{m k\ell}\Delta_{kn})=0, \label{condition-1bar}
\end{eqnarray}
for $\bar\epsilon$ invariance.
These conditions can be combined into a simpler form 
\begin{equation}
M_{\ell mn} = M_{\ell nm}=M_{n m\ell}=N_{\ell mn}= N_{\ell nm}=N_{n m\ell}, 
\label{condition-2-0}
\end{equation}
\begin{equation}
\sum_{k}(\Delta_{ k \ell}M_{kmn} +M_{\ell k n}\Delta_{km} + M_{\ell mk}\Delta_{kn})=0. 
\label{condition-2}
\end{equation}
\noindent
With  (\ref{symmetric diff}), the relation (\ref{condition-2}) implies 
a Leibniz rule on lattice
\begin{equation}
\sum_{k}\Delta_{\ell k}\sum_{m,n}M_{kmn}\phi_m\psi_n=
\sum_{k,n}M_{\ell kn}(\Delta\phi)_k\psi_n+
\sum_{m,k}M_{\ell mk}\phi_m(\Delta\psi)_k,
\label{LR}
\end{equation}
\noindent
where $\phi,\psi$ are any lattice fields.

Before  going into the further analysis, we describe  the translational invariant and local  difference operator  and  product-coefficients as  holomorphic functions\cite{KSS}. 
From translational invariance, these objects depend only on the difference of site indices, so we denote conveniently
\begin{equation}
\Delta(k) \equiv \Delta_{m\: m-k}, ~~  M(k,\ell)\equiv M_{m\: m-k\: m-\ell}.
\end{equation}
\noindent
By transforming them to the momentum representation ($w-$expression), the complex functions
\begin{equation}
\hat{\Delta}(w) \equiv \sum_{n=-\infty}^{\infty} w^n \Delta(n),~~
\hat{M}(w,z) \equiv \sum_{m,n=-\infty}^{\infty} w^mz^n M(m,n) = \hat{M}(z,w)
\label{w-expression}
\end{equation}
are defined.  Substituting $w=\exp(ipa),~z=\exp(iqa)$,
the expression (\ref{w-expression}) exactly corresponds to its Fourier expansion 
 and $p,q$ become momentum variables when $p,q$ are real.
 The locality of the difference operator  is 
exactly equivalent to the holomorphism of the $w$-expression,
and the holomorphic domain ${\cal D}$ is an annulus in the complex $w$-plane which includes  a unit circle 
around the origin.   For  the product $\hat{M}(w,z)$, the domain  is  
${\cal D} \otimes {\cal D}$ as  a two-variables function. 
Namely, we proved that  the locality, {\it i.e.} long distance behaviors 
\begin{equation}
|\Delta_{m\,n}| \leq C\exp{(-K_0|m-n|)},~~|M_{\ell m n}| \le C_1\exp{(-K_1|\ell - m|-K_2|\ell -n|)} 
\end{equation}
\noindent
with $C,~C_1,~K_0,~K_1,~K_2 >0$ are equivalent to the holomorphism of $\hat{\Delta}(w)$ and $\hat{M}(w,z)$\cite{KSS}. 
From the second relation of (\ref{diff}), the correspondence $\hat{\Delta}$ has the property, 
\begin{equation}
\hat{\Delta}(1)=0 .
\end{equation}
\noindent
Turning the relation (\ref{condition-2}) into the $w$-expression,  we can find 
\begin{equation}
\hat{M}(w,z)\hat{\Delta}(\frac{1}{wz})+\hat{M}(w,z)\hat{\Delta}(w)+
\hat{M}(w,z)\hat{\Delta}(z)=0. 
\label{LR3}
\end{equation}
\noindent
The only holomorphic solution around $w=z=1$ of (\ref{LR3}) is 
$\hat{\Delta}(w)\sim \log w$  {\it i.e.} a SLAC-type difference operator besides 
a trivial solution with $\hat{M}(w,z)=0$. 
The SLAC-type operator is, however, not  holomorphic in the annulus ${\cal D}$.  
Thus there is no set of $\Delta$ and $M(\ne 0)$ with  
translational invariance  and locality  that satisfies (\ref{condition-2}). This is an essence of
the no-go theorem for the Leibniz rule on lattice proved in our previous work\cite{KSS}.
So the lack of lattice Leibniz rule is an obstruction 
for lattice supersymmetry with  interaction.

To overcome the above problem, we demand only a subset of supersymmetry.
For example, if we impose the invariance only for 
 $\epsilon$ in (\ref{SUSY-1}),  then
 we have  the half of the condition, {\it i.e.} (\ref{condition-1}) without (\ref{condition-1bar}). Thus, instead of (\ref{condition-2-0}) and (\ref{condition-2}), we obtain
\begin{equation}
M_{\ell mn}=M_{\ell nm}=N_{nm\ell}=N_{mn\ell},
\label{symmetric Leibniz 0} 
\end{equation}
\begin{equation}
\sum_k (\Delta_{k\ell}M_{kmn}+M_{kn\ell}\Delta_{km}+M_{k\ell m}\Delta_{kn})=0 .
\label{symmetric Leibniz} 
\end{equation}
We call  (\ref{symmetric Leibniz}) as {\it cyclic  Leibniz rule} (CLR). 
This relation cannot be factorized, like (\ref{LR3}), into a separate condition for a difference operator, but must be solved by the combination of a difference  operator and  a product. 
It is  worth noting that   the symmetry in  the product  $M_{k\ell m}$ of ({\ref{condition-2-0}) is different from  
 that of  (\ref{symmetric Leibniz 0}); the former is totally-symmetric for the three indices 
 and the latter is only symmetric for the second and the third indices.   
Although  the conventional  LR (\ref{condition-2}) seems equivalent to the CLR (\ref{symmetric Leibniz})  in the naive continuum limit because  the  limit implies $M_{k\ell m} \rightarrow \delta_{k\ell} \delta_{km} $, 
they clearly have different property for the finite lattice constant. 
In contrast to (\ref{LR}), the CLR can be expressed as 
\begin{eqnarray} 
\sum_{k}\Delta_{\ell k}\sum_{m,n}M_{kmn}\phi_m\psi_n
= \sum_{n,k}   M_{kn\ell}(\Delta \phi)_k\psi_{n} +   \sum_{k,m}  M_{k \ell m}   \phi_{m}(\Delta \psi)_k ,
\end{eqnarray}
\noindent
whose indices are indeed summed in a different way from (\ref{LR}).
Although the CLR is found in constructing a specific model with cubic interaction,
it will turn out to be very useful for the analysis of more general
supersymmetric theories in later sections.
Instead of the usual Leibniz rule (\ref{condition-2}), we can adopt the CLR 
as the key  rule for the exact lattice supersymmetry.
As seen later, we can construct a superfield formalism and find a Nicolai map 
with vanishing surface terms in interacting cases. 
 This framework with the exact symmetry  are  a real advantage of our approach .   

In $w$-represenation,  
the cyclic  Leibniz rule (\ref{symmetric Leibniz}) can be expressed as 
\begin{equation}
\hat{M}(w,z)\hat{\Delta}(\frac{1}{wz})+\hat{M}(z,\frac{1}{wz})\hat{\Delta}(w)+\hat{M}(\frac{1}{wz},w)\hat{\Delta}(z)=0 .
\label{hat-CLR}
\end{equation}
\noindent
We give here an example solution for 
(\ref{hat-CLR})
keeping translational invariance  and  locality 
\begin{equation}
\hat{M}(w,z) =\frac{1}{6}\Big(2wz+wz^{-1}+zw^{-1}+2(wz)^{-1}\Big), ~~\hat{\Delta}(w)=\frac{w-w^{-1}}{2} .
\end{equation}
\noindent
This corresponds to a real space expression
\begin{eqnarray}
M_{\ell mn} &=&\frac{1}{6}(2\delta_{\ell, m-1} \delta_{\ell, n-1} 
+ \delta_{\ell, n-1} \delta_{\ell, m+1}+ \delta_{\ell, n+1}\delta_{\ell, m-1} 
+ 2 \delta_{\ell, m+1}\delta_{\ell, n+1}), \nonumber \\
\Delta_{mn} &=& \frac{1}{2}(\delta_{m, n-1} - \delta_{m, n+1}).
\label{eg-1}
\end{eqnarray}
\nonumber
This simple example has doublers which can be resolved by a supersymmetric Wilson term or a 
supersymmetric overlap-type mass. 
We shall see these terms explicitly in the subsection \ref{construction}.
More generic other solutions to (\ref{hat-CLR}) are given in Appendix \ref{appendix_solution}.

\vspace{5mm}
%
%
%
\section{Analysis of supersymmetric lattice quantum mechanics}
%
%
In the previous section,  we have found the important relation, the CLR.  Its solution consists of a local difference operator and a symmetric field product   smeared over the lattice sites keeping locality.  
In this section, we explicitly construct 
general supersymmetric quantum mechanics using the CLR as a guiding principle.
And we will analyze the model by localization technique.

\subsection{Construction of supersymmetric action}\label{construction}
We saw in the previous section that a cubic action can be half-supersymmetric if the product satisfies the CLR. This can be immediately extended to other order terms including quadratic, quartic or higher.
An important point is that the CLR for each order guarantees the invariance of each order term independently. In other words, the coupling constant for each order term can be taken independently. This naturally leads us to the superfield formalism with half-supersymmetry to construct generic action.
Also we are able to utilize this property to make an exact calculation with localization technique which will be discussed in the next subsection.

Let us first define a 2-body product of lattice fields using the coefficient $M_{\ell mn}$
\begin{equation}
 \{ \phi, \psi  \}_{\ell} \equiv  \sum_{m n} M_{\ell m n}\phi_{m}\psi_n =
  (-1)^{\epsilon_{\phi} \epsilon_{\psi}}\{ \psi, \phi  \}_{\ell}, 
\label{smeared s-product}
\end{equation}
where ${\epsilon_{\phi}}$ and ${\epsilon_{\psi}}$ are the Grassmann-parity for $\phi$ and $\psi$.
We  refer  this product as {\it a smeared symmetric product}, because this is a symmetric product smeared over lattice sites. The locality of the product is guaranteed by the holomorphic property of its $w$-expression $\hat M(z,w)$.   
We also define an inner product of two lattice fields
\begin{equation}
(\phi, \psi) \equiv  \sum_n \phi_n  \psi_n.
\label{inner product}
\end{equation}
Then a simple 3-body summed product of lattice fields  is expressed as 
\begin{equation}
(\phi, \{ \psi, \chi  \}) \equiv  \sum_{\ell m n} M_{\ell m n}\phi_{\ell}\psi_m \chi_n. 
\label{3-body inner product}
\end{equation}
With these notations, as well as (\ref{symmetric diff}), (\ref{symmetric Leibniz}), the CLR can be rewritten into a compact form 
\begin{equation}
(\Delta\phi,\{\psi,\chi\} )+ (\Delta\psi,\{\chi,\phi\} )+(\Delta\chi,\{\phi,\psi\} )=0. 
\label{2-product-0}
\end{equation}
\noindent
In the case of three same  fields,  (\ref{2-product-0}) is reduced to 
\begin{equation}
(\Delta\phi,\{\phi,\phi\} )=0.
\label{2-product-1}
\end{equation}
\noindent
The expression (\ref{2-product-1}) is equivalent to (\ref{2-product-0})
if we take the symmetry of the product $M$ into account.  These simple expressions are useful in  
further SUSY analysis.

Now we construct various terms in the action using superfield formalism.
As stated before, the kinetic term is fully supersymmetric with $\epsilon$ and $\bar\epsilon$, while the interaction term is half-supersymmetric with, say $\epsilon$. Therefore we can use ${\cal N}=2$ superspace for the former, but 
must use ${\cal N}=1$ for the latter.

An ${\cal N=}2$ supermultiplet for a superspace $(n,\theta,\bar{\theta})$
\begin{equation}
{\Xi}_n(\theta,\bar{\theta})= \Phi_n(\theta)- \bar{\theta}\Psi_n(\theta)
\label{multiplet-1}
\end{equation}
\noindent
is decomposed into two ${\cal N=}1$ multiplets,  
\begin{equation}
\Phi_n(\theta)= \phi_n + \theta \bar{\psi}_n 
\label{multiplet-2}
\end{equation}
\noindent
which is Grassmann even and
\begin{equation}
 \Psi_n(\theta)= \psi_n + \theta F_n 
 \label{multiplet-3}
\end{equation}
\noindent
which is Grassmann odd.

A kinetic term of the action can be written as 
\begin{equation}
S_0 = \frac{1}{2}\int d\bar{\theta} d\theta (\bar{D}{\Xi}(\theta,\bar{\theta}),D{\Xi}
(\theta,\bar{\theta}))  ,
\label{free-action}
\end{equation}
\noindent
where  supercovariant difference operators are defined as 
\begin{equation}
\bar{D}_{mn} \equiv  -i \frac{\partial}{\partial \theta} \delta_{mn} - \bar{\theta} \Delta_{mn} , \  \
 D_{mn} \equiv  -i \frac{\partial}{\partial \bar{\theta}} \delta_{mn} - \theta \Delta_{mn} .
\label{supercovariant}
\end{equation}

To formulate mass term and Wilson term supersymmetrically, we consider a generic bilinear form 
\begin{equation}
\label{2-body}
(\phi, G \psi )  \equiv \sum_{k\ell}\phi_k G_{k\ell} \psi_\ell ,
\end{equation}
where $G_{mn}$ is a certain operator on a lattice.
The translational invariance implies that $G$ is a function of the difference of its indices $G_{mn}=G(m-n)$, and satisfies $\Delta G = G \Delta$ with translationally-invariant difference operator $\Delta$. Thus we have
\begin{equation}
(\Delta\phi, G\psi )=-(\phi, G\Delta \psi) ,
\label{DeltaG}
\end{equation}
with
$\Delta^T=-\Delta$.
Then
\begin{equation}
S_2= i \int d\theta \, (\Psi(\theta), G \Phi(\theta)  )
 \label{bilinear}
\end{equation}
gives a supersymmetric quadratic action invariant under $\epsilon$-transformation.
For example, a simple mass term with Wilson term is given by choosing
\begin{equation}
\hat{G}(w)\equiv \sum_n w^nG(n) = im+ ir\frac{2-w-w^{-1}}{2}
\end{equation}
where $m$ is a bare mass and $r$ is a standard Wilson term parameter.
For more generic doubler-suppressing term, we can write
\begin{equation}
\hat{G}(w) = im+i\hat H(w,M) , 
\end{equation}
\noindent
where $\hat H(w,M)$ is a holomorphic function in ${\cal D}$ with a parameter $M>0$ and corresponds to $w$-expression of a translationally invariant operator $H_{mn}(M)$, {\it i.e.}~$\hat H(w,M)=\sum_mw^mH_{\ell\,\ell+m}(M)$.
We further impose that $H$ is symmetric, {\it i.e.}~$H^T=H$ or $\hat H(w)=\hat H(1/w)$, throughout this paper.
If we need an overlap-type action, we choose $\hat H$ to have the poles at $w=0,\infty$ and the branch points at $w=-1/M, -M$ with $M>1$, as well as the property $\hat H(w=1,M)=0$.
The explicit form for the overlap-type term\cite{overlap-1, overlap-2}  is given as
\begin{equation}
\frac{\hat H(w,M)}{M+1}= 1- \frac{(M+\frac{w+w^{-1}}{2})}{\sqrt{-(\frac{w-w^{-1}}{2})^2+(M+\frac{w+w^{-1}}{2})^2}}
= 1-\frac{(M+\frac{w+w^{-1}}{2})}{\sqrt{(M+w)(M+w^{-1})}} ,
\end{equation}
\noindent

Note that, together with an explicit solution for the CLR
\begin{equation}
\hat\Delta(w)=\frac{(M+1)(w-w^{-1})}{2\sqrt{(M+w)(M+w^{-1})}} ,
\end{equation}
\begin{equation}
\hat{M}(w,z)=\frac{1}{6(M+1)}\sqrt{(M+w)(M+w^{-1})}\Big(2wz+wz^{-1}+zw^{-1}+2(wz)^{-1}\Big),
\end{equation}
the above $H$ satisfies an analog of the Ginsparg-Wilson relation\cite{Ginsparg-Wilson,Neuberger}, 
\begin{equation}
\hat\Delta(w)\hat\Delta(w^{-1}) + \hat H(w,M)\hat H(w^{-1},M)=  (M+1)\Big(\hat H(w,M)+\hat H(w^{-1},M)\Big), 
\end{equation}
\noindent
where  the Dirac operator  is given by   
$\hat{D}_{Dirac}(w) = i\hat\Delta(w)+i\hat H(w,M)$.

The quadratic terms including mass and doubler-suppressing term and the cubic interaction term of the supersymmetric action are now obtained as
\begin{equation}
S_2+S_{{\rm int}} = i \int d\theta \Big(  (\Psi(\theta), G \Phi(\theta)  )
+ g_2(\Psi(\theta),\{ \Phi(\theta), \Phi(\theta)  \} )
 \Big) 
 \label{int-action}
\end{equation}
\noindent
with a coupling constant $g_2$ (here subscript 2 stands for 2-body product).
The action (\ref{int-action}) is clearly 
invariant under SUSY transformation
\begin{eqnarray}
 \delta \phi_n&=& \epsilon \bar{\psi}_n,\nonumber \\
 \delta \psi_n&=& \epsilon (i(\Delta\phi)_n+F_n), \nonumber \\
 \delta \bar{\psi}_n&=& 0,\nonumber \\
 \delta F_n&=&  -i\epsilon  (\Delta\bar{\psi})_n  .
\label{1/2SUSY}
\end{eqnarray}
\noindent
This transformation can be realized on the superfield as follows,
\begin{equation}
{\delta} \Xi_n(\theta,\bar{\theta}) = \sum_m
\epsilon \Big(\frac{\partial}{\partial \theta}\delta_{nm}
+ i\bar{\theta} \Delta_{nm}\Big)
\Xi_m(\theta,\bar{\theta}) =\epsilon \Big(\frac{\partial}{\partial \theta} \Phi_n(\theta)
+\bar{\theta}\frac{\partial}{\partial \theta}\Psi_n(\theta) +i \bar{\theta} (\Delta\Phi)_n(\theta)\Big)  ,
\label{N=1-1}
\end{equation}
\noindent
or equivalently on the ${\cal N}=1 $ superfields
\begin{eqnarray}
\delta\Phi_n(\theta)&=&\epsilon \frac{\partial}{\partial \theta}\Phi_n(\theta),  \nonumber \\
\delta \Psi_n(\theta) &=& \epsilon\frac{\partial}{\partial \theta}\Psi_n(\theta)+i\epsilon (\Delta \Phi)_n(\theta) . 
\label{N=1-2}
\end{eqnarray}

Thus the total action
\begin{eqnarray}
S_{{\rm t}}&=& S_0+S_2+S_{{\rm int}}\nonumber\\[5pt]
&=& \frac{1}{2}(\Delta \phi,  \Delta \phi)+ i (\bar{\psi}, \Delta \psi)+
\frac{1}{2}(F,F)\nonumber\\[3pt]
&+&(F,(im+iH)\phi) +
(\bar{\psi},(im+iH)\psi)+ig_2(F,\{\phi,\phi\})-2ig_2(\psi,\{\phi,\bar{\psi}\})
\label{total action}
\end{eqnarray}
is also invariant under (\ref{N=1-1}) or (\ref{N=1-2}), 
\begin{equation}
 \delta S_{\rm t} =0,
\label{invariance}
\end{equation}
due to the relation (\ref{DeltaG}) and the CLR (\ref{2-product-1}).
It is worth emphasizing here that our formulation keeps locality; the holomorphic property of the $w$-expression of various operators and product guarantees the locality.

To extend to multi-body supersymmetric interactions, we define  the smeared   $N$-body products for a bosonic superfield
\begin{equation}
\{\Phi,\ldots, \Phi  \} _n  \equiv  \sum_{m_1,\ldots,m_{N}} M_{n;m_1,\ldots,m_{N}} \Phi_{m_1}  \cdots  \Phi_{m_{N}} .
\label{N-Product-1}
\end{equation}
\noindent
where $m_1,..,m_N$  are totally symmetric indices
(we have been omitting semicolon for $N=2$, {\it i.e.} $M_{n;m_1,m_2} = M_{nm_1m_2}$.) 

If this product satisfies the $N$-body CLR
\begin{equation}
(\Delta \Phi, \{\Phi,\ldots, \Phi \} ) =0 ,
\label{N-Product-2}
\end{equation}
then the action 
\begin{equation}
\int d\theta\,i g_{N} \left( \Psi, \{\Phi,\ldots,\Phi \} \right)
\label{N-Product-interaction}
\end{equation}
%
is invariant under the supersymmetry transformations (\ref{N=1-2}) in exactly same way as the 2-body case.
It should be noticed that the $R$-symmetry
%
\begin{eqnarray}
\theta &\to& e^{i\alpha} \theta\,,\nonumber\\
\Phi &\to& \Phi\,,\nonumber\\
\Psi &\to& e^{i\alpha}\Psi\,,
\label{Rsymmetry}
\end{eqnarray}
%
allows only the type of interaction terms (\ref{N-Product-interaction}),
which consists of $\Phi$ and $\Psi$
and is necessary  for constructing supersymmetric 
$2N$-body interactions for scalar fields with a coupling constant $g_N$. 
Also note that $N$-body product satisfying the CLR can be constructed in terms of 2-body products which will be discussed in Appendix~B.

Before closing this subsection, it may be helpful to understand the exact SUSY invariance in an alternative way.
We write the total action $S_{\rm t}$  as a sum of kinetic term $S_0$, bare mass term $S_m$, doubler-suppressing term $S_{\rm ds}$ and interaction term $S_{\rm int}$
\begin{equation}
S_{\it t}=S_0+S_m+S_{\rm{ds}}+S_{{\rm int}}.
\label{total action'}
\end{equation}
Then each term can be expressed as a supersymmetric transform of something
\begin{eqnarray*}
 S_0&=& \frac{1}{2}\delta'(\psi, (-i\Delta \phi +F)), \\
 S_m&=& im\delta'(\psi, \phi ),   \\
 S_{\rm{ds}}& = & i\delta'(\psi, H \phi ),  
 \end{eqnarray*}
 \begin{equation}
 S_{{\rm int}}=S_{3}+S_{4}+ \cdots =ig_2\delta'(\psi, \{\phi, \phi\} ) 
 + ig_3 \delta'(\psi, \{\phi, \phi, \phi\} ) +\cdots ,
\end{equation}
\noindent
with coupling constants $g_2,g_3,\cdots$. Here we denote the transformation without parameter by $\delta'$, {\it i.e.}  $\delta = \epsilon \delta'$,
or explicitly
\begin{eqnarray}
\delta'\phi_n&=&\bar{\psi}_n,  \nonumber \\
\delta' \psi_n &=& i(\Delta \phi)_n+F_n,  \nonumber \\ 
\delta' \bar{\psi}_n &=&0, \nonumber \\
\delta'F_n &=& -i(\Delta  \bar{\psi})_n .
\label{N=1'-2}
\end{eqnarray}
Due to the nilpotencey of the transformation $\delta'^2=0$,
the invariance of each term is manifest,
\begin{equation}
\delta' S_{0}=\delta' S_m=\delta'S_{\rm{ds}}=\delta' S_{{\rm int}}=0.
\end{equation}

\subsection{Localization technique and calculation of Witten index}

With the action (\ref{total action'}), we can find a Nicolai map for our system, and apply the localization technique\cite{local-1,local-2} to calculate the Witten index (the partition function with periodic boundary condition for all variables).

Before going into the detail, it would be useful to compare with the previous attempts in the existing literature.
In the references \cite{Kaplan-Catterall,LSUSY5} the partition function has been calculated in the ultra local
limit, where the $N$-site partition 
function $Z_{N}$ 
reduces to $N$ copies of the one-site one, {\it i.e.} $Z_{N} = (Z_{1})^{N}$.
This result indicates that every degrees of freedom may contribute
to the Witten index that is quite different from the usual understanding of its topological nature. Also it depends on whether the number of lattice sites is even  or odd for $Z_1=-1$ case of which we could not understand the validity.
On the other hand, our lattice formulation is suitable to use
the localization technique as will be seen below.
It will turn out that the Witten index is determined by only zero modes
and all fluctuation modes are irrelevant in it.

Now let us begin by defining the Nicolai map 
\begin{equation}
\xi_n=-(\Delta \phi)_n +  m \phi_n + (H\phi)_n +g_2\{ \phi,  \phi \}_n + g_3 \{ \phi,  \phi, \phi \}_n + \cdots.
\label{Nicolai_map}
\end{equation}
\noindent
With this, the Witten index can be  calculated  as 
\begin{eqnarray}
Z=\int D\phi  D\bar{\psi}D\psi DF \,e^{-S_{\rm t}}&=&\int D\xi D\bar{\psi}D\psi 
|\det \eta|^{-1} \,e^{-\sum_n \frac{1}{2}\xi_n \xi_n-i\sum_{kl}\bar{\psi}_{l}\eta_{lk}\psi_{k}}   \nonumber \\
 & =  &    \int D\xi 
|\det \eta|^{-1}  \det \eta~
e^{-\sum_n \frac{1}{2}\xi_n \xi_n} ,
\label{path-integral-1}
\end{eqnarray}
\noindent
where all variables follow a periodic boundary condition and 
\begin{equation}
\eta_{\ell k} \equiv \frac{\partial \xi_k}{\partial \phi_{\ell}}=\Delta_{\ell k} + m\delta_{\ell k}+{(H)}_{\ell k}  
+g_2\frac{\partial }{\partial \phi_{\ell}}\{\phi, \phi  \}_{k} + 
g_3\frac{\partial }{\partial \phi_{\ell}}\{ \phi,  \phi,  \phi \}_{k} + \cdots .
\end{equation}
\noindent
The prefactor of (\ref{path-integral-1})
\begin{equation}
|\det \eta|^{-1}\det \eta  ,
\end{equation}
\noindent
is not generally a constant but $\det\eta$ can vanish for certain cases. 
Also the path of the integral  (\ref{path-integral-1}) is not definite.      
Therefore we must carry out more detailed analysis. 

In order to control the infrared behavior, we restrict a size of the system to finite $N$. Also we assume the highest power of the potential in (\ref{Nicolai_map}) is $p$, for definiteness.
We rescale our kinetic and doubler-suppressing term with parameters $t$ and $t_{\rm ds}$ respectively, 
\begin{equation}
S_0 \rightarrow tS_0 ,~~S_{\rm{ds}} \rightarrow t_{\rm{ds}}S_{\rm{ds}}.
\end{equation}
\noindent
Note that $t_{\rm{ds}}$ looks like a Wilson term parameter for Wilson fermion case. Then we integrate out $F_n$ and obtain  
 a $t$-scaled partition function, 
\begin{eqnarray}
 Z (t,t_{\rm{ds}})&\equiv& \int D\phi D\bar{\psi}  D\psi DF\,
 e^{-S(t,t_{\rm{ds}})} \nonumber \\
  &\equiv&  \int D\phi D\bar{\psi}  D\psi DF\,
  e^{-tS_0-t_{\rm{ds}}S_{\rm{ds}} -S_m -S_{{\rm int}}} \nonumber
  \\
 & = &t^{-N/2}\int D\phi D\bar{\psi}  D\psi~ \nonumber \\ 
 &\times& e^{-\big(tS^b_0   
  + t^2_{\rm{ds}}t^{-1} S^b_{\rm{ds}} + t^{-1} S^b_{m+{\rm int}}
  + t_{\rm{ds}}t^{-1}S^b_{\rm cross}\big)
- \big(tS^f_0+  t_{\rm{ds}}S^f_{\rm{ds}}+S^f_{m + {\rm int}}\big)} . 
\label{t-scaled Z}
\end{eqnarray}
Here we denote  $S^b_0, S^b_m, S^b_{\rm{ds}}$ and $S^b_{{\rm int}}$ as bosonic (without fermion) parts of kinetic,  mass, doubler-suppressing   and interaction terms, respectively.  
$S^f_0, S^f_m, S^f_{\rm{ds}}$ and $S^f_{{\rm int}}$  are  fermionic (with fermion) counterparts.
\begin{table}[hb]
\caption{Bosonic and fermionic parts of the action after integrating of $F_n$}
\begin{center}
$$
\begin{array}{|c|c|c|}
\hline
S_0^b & S^b_{\rm ds} & S^b_{m+{\rm int}}\rule[-6pt]{0pt}{18pt}\\ \hline \hline
\frac{1}{2}(\Delta \phi, \Delta \phi)  & \frac{1}{2}(\phi, H^2 \phi) & 
\frac{1}{2}(m\phi+g_2\{\phi,\phi\}+\cdots, m\phi+g_2\{\phi,\phi\}+\cdots) \rule[-6pt]{0pt}{18pt}\\ \hline
\end{array}
$$
$$
\begin{array}{|c|c|c|} \hline
S_0^f & S^f_{\rm ds} & S^f_{m+{\rm int}}\rule[-6pt]{0pt}{18pt}\\ \hline \hline
i(\bar{\psi}, \Delta \psi) &  i(\bar{\psi},H \psi) & im(\bar{\psi},  \psi)-i(\psi, 2g_2\{ \phi , \bar{\psi}\}+\cdots) \rule[-6pt]{0pt}{18pt}\\ \hline
\end{array}
$$
\end{center}
\label{table2}
\end{table}
We summarize each part of our action in Table~\ref{table2}.
In addition, the cross terms $S^b_{\rm cross}$ for doubler-suppressing term and mass plus interaction terms are defined as 
\begin{equation}
S^b_{\rm cross} \equiv (H\phi\,,\,m\phi+g_2\{\phi,\phi\}+\cdots).
\end{equation}
 
 It is clear that this $Z(t,t_{\rm{ds}})$
 is $t$-independent,
because before $F_n$-integration one can show
\begin{eqnarray}
\frac{\partial}{\partial t}Z(t,t_{\rm{ds}})&=&-\int D\phi D\bar{\psi}  D\psi DF ~S_0~e^{-tS_0 - S_m - t_{\rm{ds}}S_{\rm{ds}} -S_{{\rm int}}} \nonumber \\ 
 &=& -\int D\phi D\bar{\psi}  D\psi  DF\,\delta' (Xe^{-tS_0 -S_m -t_{\rm{ds}}S_{\rm{ds}} -S_{{\rm int}}}) =0 ,
\end{eqnarray}
\begin{eqnarray}
\frac{\partial}{\partial t_{\rm{ds}}}Z(t,t_{\rm{ds}})&=&-\int D\phi D\bar{\psi}  D\psi DF ~S_{\rm{ds}}~e^{-tS_0 - S_m - t_{\rm{ds}}S_{\rm{ds}} -S_{{\rm int}}} \nonumber \\ 
 &=& -\int D\phi D\bar{\psi}  D\psi  DF\,\delta' (Ye^{-tS_0 -S_m -t_{\rm{ds}}S_{\rm{ds}} -S_{{\rm int}}}) =0 ,
\end{eqnarray}
\noindent
where $X\equiv\frac{1}{2}\Big(\psi, -i\Delta \phi + F\Big),~Y\equiv \Big(\psi,H\phi  \Big)$.
Here we used the invariance of the path-integral measure under $\delta'$ and $\delta' (S(t,t_{\rm{ds}}))=0$.

In the following calculations of the Witten index, we take   
two typical limiting cases.  One is analogous to the continuum theory in which  important modes are only zero-modes. The second case  is   peculiar to 
lattice theories, which has doubling modes in addition to zero-modes.\footnote{Hereafter we assume the number of sites $N$ is even for simplicity. 
There is no doubler mode for odd $N$, so that it gives the same result
 as Case~(i) below. } 
Although  the two cases are physically and mathematically different,  
after using localization technique, we can obtain    
the exactly same results. 
We first decompose each field into zero mode (script $0$), doubler mode (script $d$) and the other modes:
\begin{eqnarray}
\phi_n&=&\frac{\phi_0}{\sqrt{N}}+ \frac{(-1)^n}{\sqrt{N}}\phi_d +a_n ,  \nonumber \\
\psi_n&=&\frac{\psi_0}{\sqrt{N}} + \frac{(-1)^n}{\sqrt{N}}\psi_d +\chi_n, \nonumber \\
\bar{\psi}_n&=&\frac{\bar{\psi}_0}{\sqrt{N}} +
 \frac{(-1)^n}{\sqrt{N}}\bar{\psi}_d +\bar{\chi}_n 
\end{eqnarray}
Then we rescale each part with $t$ or $t_{\rm ds}$ in such a way that the only important mode survives in the infinitely scaling limit.

\vspace{0.5cm}

\underline{Case (i)}~~~We make the scaling of each field as
\begin{eqnarray}
\phi_n&=&\frac{\phi_0}{\sqrt{N}}+ \frac{(-1)^n}{\sqrt{N}\sqrt{t_{\rm{ds}}}}\phi_d +\frac{1}{\sqrt{t}}a_n ,  \nonumber \\
\psi_n&=&\frac{\psi_0}{\sqrt{N}} + \frac{(-1)^n}{\sqrt{N}\sqrt{t_{\rm{ds}}}}\psi_d +\frac{1}{\sqrt{t}}\chi_n, \nonumber \\
\bar{\psi}_n&=&\frac{\bar{\psi}_0}{\sqrt{N}} +
 \frac{(-1)^n}{\sqrt{N}\sqrt{t_{\rm{ds}}}}\bar{\psi}_d +\frac{1}{\sqrt{t}}\bar{\chi}_n ,
\end{eqnarray}
then we take $t,t_{\rm{ds}} \rightarrow \infty$  with a fixed $t/t_{\rm{ds}}$. 
This case is similar to  the continuum theory, because
the real zero mode (not doubler) is only $t,t_{\rm{ds}}$-unscaled. 
A scalar zero mode  $\phi_0$ contributes to 
$t^{-1}S^b_{m+{\rm int}}+S^f_{m+{\rm int}}$
and a scalar doubler mode $\phi_d$ contributes to 
\begin{equation}
\frac{t^2_{\rm{ds}}}{t}S^b_{\rm{ds}}=\frac{t_{\rm{ds}}}{tN}\sum_{m}\Big(\sum_n H_{mn} (-1)^n \phi_d\Big)^2
=\frac{t_{\rm{ds}}}{t}\big(\hat{H}(-1)\big)^2\phi_d^2,  
\end{equation}
\noindent
where $\hat{H}(w)\equiv \sum_n w^n H(n)$ for translational invariant $H_{mn}=H(m-n)$.
On the other hand, the fermion zero modes contribute to 
$S^f_{m+{\rm int}}$,
and   a fermion doubler mode contributes to 
\begin{equation}
t_{\rm{ds}}S^f_{\rm{ds}}=i \frac{1}{N}\sum_{mn}(-1)^m\bar{\psi}_d H_{mn} (-1)^n \psi_d =
 i \hat{H}(-1)\bar{\psi}_d \psi_d   . 
\end{equation}
\noindent
Any non-zero modes  can be integrated out in the  limit  $t,t_{\rm{ds}} \rightarrow  \infty$  with fixed $t/t_{\rm{ds}} $, and 
the $t$ factor of the partition function (\ref{t-scaled Z}) is exactly canceled.
$(t_{\rm{ds}}/t) S^b_{\rm{cross}}$ terms in the action  
vanish  in the limit.

Therefore
\begin{eqnarray}
Z(t,t_{\rm{ds}}) &=&\frac{\sqrt{t_{\rm{ds}}}}{t}\int d\phi_0 d\phi_d  
d\bar{\psi}_0d\psi_0 d\bar{\psi}_d d\psi_d  ~e^{-(t_{\rm{ds}}^2t^{-1}S^b_{\rm{ds}}+t^{-1}S^b_{m+{\rm int}})-(t_{\rm{ds}}S^f_{\rm{ds}}+
S^f_{m+{\rm int}})}     \nonumber \\
 &=&\frac{\sqrt{t_{\rm{ds}}}}{t}  \int d\phi_d \bar{\psi}_d \psi_d ~e^{-t_{\rm{ds}}^2t^{-1}S^b_{\rm{ds}}-t_{\rm{ds}}S^f_{\rm{ds}}} 
 \int d\phi_0 d\bar{\psi}_0d\psi_0 ~e^{-t^{-1}S^b_{m+{\rm int}}-S^f_{m+{\rm int}}}
\nonumber \\
 & = & \frac{\sqrt{t_{\rm{ds}}}}{t} \sqrt{\frac{t}{t_{\rm{ds}}
 \big(\hat{H}(-1)\big)^{2}}} \,\hat{H}(-1) \sqrt{t}
 \int d\xi |\eta_0|^{-1}\eta_0 ~e^{-\frac{1}{2}\xi^2} ,
 \label{Z-i}
 \end{eqnarray}
\noindent
where 
\begin{eqnarray}
\xi(\phi_0)&\equiv&\frac{1}{\sqrt{t}}\left(m\phi_0+ \frac{g_2}{\sqrt{N}} \hat{M}(1,1)\phi_0^2  +\frac{g_3}{N} \hat{M}(1,1,1)\phi_0^3 + \cdots 
+\frac{g_{p}}{\sqrt{N}^{p-1}} \hat{M}(1,\cdots,1)\phi_0^{p}\right) , \nonumber \\
\eta_0(\phi_0) &\equiv& m+ \frac{2g_2}{\sqrt{N}} \hat{M}(1,1)\phi_0+ \frac{3g_3}{N} \hat{M}(1,1,1)\phi_0^2 + \cdots + 
\frac{pg_{p}}{\sqrt{N}^{p-1}} \hat{M}(1,\cdots,1)\phi_0^{p-1} \qquad.
\end{eqnarray}
with $w$-expression of translaionally invariant multi-body product 
\begin{equation}
 \hat M(w_1,\cdots,w_{n-1})=\sum_{m_1,\cdots,m_{n-1}}
 w_1^{m_1}\cdots w_{n-1}^{m_{n-1}}
 M_{\ell;\ell+m_1,\cdots,\ell+m_{n-1}} .
 \end{equation}
The final expression (\ref{Z-i}) results in $Z(t,t_{\rm{ds}})=0$ for $p=$even and  $Z(t,t_{\rm{ds}})=\pm 1$ for $p=$odd (its sign depends on the asymptotic behavior of the potential).

\vspace{0.5 cm}

\underline{Case (ii)}~~~An alternative way of rescaling of fields is given by 
\begin{eqnarray}
\phi_n&=&\frac{1}{\sqrt{N}}\phi_0 + \frac{(-1)^n}{\sqrt{N}}\phi_d +\frac{1}{\sqrt{t}}a_n , \nonumber \\ 
\psi_n&=&\frac{1}{\sqrt{N}}\psi_0 + \frac{(-1)^n}{\sqrt{N}}\psi_d +\frac{1}{\sqrt{t}}\chi_n, \nonumber \\
\bar{\psi}_n&=&\frac{1}{\sqrt{N}}\bar{\psi}_0+ \frac{(-1)^n}{\sqrt{N}}\bar{\psi}_d +\frac{1}{\sqrt{t}}\bar{\chi}_n.
\end{eqnarray}
Then we take the limit $t \rightarrow \infty$ with a fixed $t_{\rm{ds}}$.
In this case not only zero mode but also doubler mode are $t$-unscaled. 
A $\phi_0$  contributes to 
\begin{equation}
t^{-1}S^b_{m+{\rm int}} ,  ~~S^f_{m+{\rm int}}, 
\end{equation}
\noindent
and a $\phi_d$  contributes to 
\begin{equation}
t_{\rm{ds}}^2 t^{-1}S^b_{\rm{ds}},~~        t^{-1}S^b_{m+{\rm {\rm int}}} ,  ~~t_{\rm{ds}}t^{-1}S^b_{\rm{cross}}, ~~S^f_{m+{\rm int}}.
\end{equation}
\noindent
Fermion zero modes $\psi_0, \bar\psi_0$ contribute to 
\begin{equation}
S^f_{m+{\rm {\rm int}}},
\end{equation}
\noindent
and fermion doubler modes $\psi_d, \bar\psi_d$ contribute to 
\begin{equation}
t_{\rm{ds}}S^f_{\rm{ds}},~~S^f_{m+{\rm {\rm int}}}. 
\end{equation}
\noindent
Any non-zero modes can be integrated out similarly to the case (i)  in the limit $t \rightarrow  \infty$.

Therefore $Z$ is evaluated as
\begin{eqnarray}
Z(t,t_{\rm{ds}}) &=&\frac{1}{t}\int d\phi_0 d\phi_d d\bar{\psi}_0 d\psi_0d\bar{\psi}_d
 d\psi_d  \nonumber \\
&&~~~\times ~e^{-(t^{-1}t_{\rm{ds}}^2S^b_{\rm{ds}}+t^{-1}t_{\rm{ds}}S^b_{\rm cross} +t^{-1}S^b_{m+{\rm {\rm int}}})-(t_{\rm{ds}}S^f_{\rm{ds}}+
S^f_{m+{\rm {\rm int}}})}    \nonumber \\[7pt]
 & = &   \int d\xi_0 d\xi_d |\det \eta(\phi_0(\xi),\phi_d(\xi))|^{-1} \det \eta(\phi_0(\xi),\phi_d(\xi)) 
 e^{-\frac{1}{2}\xi_0^2-\frac{1}{2}\xi_d^2}   ,
 \label{Z-ii-2}
\end{eqnarray}
\noindent
where  
\begin{eqnarray}
\xi_\ell(\phi_0,\phi_d) &\equiv&  \frac{1}{\sqrt{t}}\left(\frac{t_{\rm{ds}}}{\sqrt{N}} \sum_n H_{\ell n} (-1)^n \phi_d
 + \frac{1}{\sqrt{N}}\sum_n m\delta_{\ell n} (\phi_0 + (-1)^n \phi_d) \right.
 \nonumber \\
&&+ \cdots + \frac{g_p}{\sqrt{N}^{p}}\sum_{n_1,\cdots,n_p} M_{\ell ;n_1,n_2,\cdots,n_{p}}\left.\rule{0pt}{18pt}
(\phi_0 + (-1)^{n_1} \phi_d) \cdots(\phi_0 + (-1)^{n_{p}} \phi_d)\right) \nonumber \\
 &\equiv& \frac{1}{\sqrt{N}}\left(\xi_0(\phi_0,\phi_d)+(-1)^\ell \xi_d(\phi_0,\phi_d)\right) ,
\end{eqnarray}
and
\begin{equation}
\eta(\phi_0,\phi_d) =\left(
\begin{array}{cc}
\displaystyle \frac{\partial\xi_0(\phi_0,\phi_d)}{\partial\phi_0} &
\displaystyle \frac{\partial\xi_0(\phi_0,\phi_d)}{\partial\phi_d} \\[11pt]
\displaystyle \frac{\partial\xi_d(\phi_0,\phi_d)}{\partial\phi_0} &
\displaystyle \frac{\partial\xi_d(\phi_0,\phi_d)}{\partial\phi_d}
\end{array}
\right).
\end{equation}
The final expression (\ref{Z-ii-2}) results in $Z(t,t_{\rm{ds}})=0$ for $p=$even and  $Z(t,t_{\rm{ds}})=\pm 1$ for $p=$odd.

\vspace{0.5cm}

 In summary, these two limiting procedures give exactly the same result,
although the doubler mode survives in the expression (\ref{Z-ii-2}) of the limit (ii). The final results reproduce known results for the supersymmetric quantum mechanics in the continuum. Also there is no subtle dependence on the number of lattice sites.

\section{Multicomponent case}  

In order to consider a higher-dimensional extension with spinorial index or a multi-flavor extension, we need a framework for multi-component fields. 
A Wilson term or an overlap-type term can be made more realistic in the model with multi-component indices.
The CLR for the multi-component case, however, is slightly modified and should be treated with care. 
In this section we will explain this in a simple situation where each field has a single index which we do not specify with either spinor or flavor at the beginning.

For the fields $(\phi_n^a, \psi_n^a, \bar\psi_n^a, F_n^a)$ with single index $a$, we consider the following supersymmetry transformation 
\begin{equation}
\begin{array}{rll}
\delta \phi^a_n &= \epsilon \bar{\psi}^a_n & = \epsilon\delta' \phi_n^a,   \nonumber \\[3pt]
\delta \psi_n^a &=  \epsilon (i(\Delta \phi \Gamma)^a_n+F^a_n)&=\epsilon \delta'\psi_n^a,   \nonumber \\[3pt]
\delta \bar{\psi}^a_n &= 0 &=\epsilon \delta' \bar{\psi}_n^a, \nonumber \\[3pt]
\delta F_n^a&= -i\epsilon (\Delta \bar{\psi}\Gamma)_n^a&=\epsilon \delta' F_n^a ,
\label{SUSY-spinor}
\end{array}
\end{equation}
where $\Gamma$ is a certain constant matrix on the index with property $\Gamma =\Gamma^T,~\Gamma^2=1$.
We can easily obtain  a free action including mass and doubler-suppressing terms invariant under the above transformation, 
\begin{eqnarray}
S_0 &=& \frac{1}{2}(\Delta\phi^a,\Delta\phi^a)
+ i (\bar{\psi}^a,(\Gamma)_{ab}\Delta \psi^b)
+\frac{1}{2} (F^a, F^a) \nonumber \\
&=& -\frac{i}{2}\delta' (\psi^a, (\Delta\phi \Gamma)^a)+\frac{1}{2}\delta' (\psi^a, F^a) ,\\
S_m &= &im(\phi^a,F^a)+ im (\bar{\psi}^a,\psi^a)= im \delta' (\phi^a, \psi^a), \\
S_{\rm{ds}}&=&i(\phi^a, H_{ab}F^b)+i(\bar{\psi}^a,H_{ab}\psi^b)=i\delta' (\phi^a, H_{ab}\psi^b),
\end{eqnarray}
\noindent
where $H_{ab}$ is a multi-component extension of $H$ in the previous sections and satisfies $\sum_m (H_{ab})_{mn}=\sum_m H_{ab}(m-n)=\hat{H}_{ab}(1)=0 $
and $((H\Gamma\Delta)_{ab})_{mn}=-((H\Gamma\Delta)_{ba})_{nm}$.
The interaction term 
\begin{equation}
S_{{\rm int}} = ig_2 \Big((F^a,\{\phi,\phi\}^a) - 2(\psi^a,\{\phi,\bar{\psi}\}^a)\Big)
= ig_2  \,\delta' (\psi^a,\{\phi,\phi\}^a) ,
\end{equation}
\noindent
is also invariant under (\ref{SUSY-spinor}) 
where we used $\displaystyle\{ \phi, \psi \}^a_k \equiv \sum_{b,c,m,n}M^{abc}_{kmn}\phi^b_m \psi^c_n$ and
\begin{equation}
\sum_a ((\Delta\phi\Gamma)^a,\{\phi,\phi\}^a)=0.
\end{equation}

The translational invariance implies that the $\Delta$ and $M$ only depend on the difference of site indices, as before, 
\begin{equation}
\Delta_{mn}=\Delta(m-n), ~M^{abc}_{kmn}=M^{abc}(k-m,k-n).
\end{equation}
\noindent
Their $w-$expressions are
\begin{equation}
\hat{\Delta}(w)\equiv \sum_mw^m\Delta(m), ~\hat{M}^{abc}(w,z)\equiv 
\sum_{mn}w^mz^nM^{abc}(m,n) .
\end{equation}
\noindent
Using  the combined quantity  
\begin{equation}
P^{abc}(w,z)\equiv \hat{\Delta}(1/(wz))\Gamma^{ad}\hat{M}^{dbc}(w,z),
\end{equation}
\noindent
the multicomponent CLR is expressed as 
\begin{equation}
\begin{array}{c}
P^{abc}(w,z)+P^{bca}(z,1/(wz))+P^{cab}(1/(wz),w)=0,\\[5pt]
P^{abc}(w,1/w)=0,\qquad P^{abc}(w,z)=P^{acb}(z,w) .
\end{array}
\label{multi-CLR}
\end{equation}
\noindent
Although solutions for the relation (\ref{multi-CLR}) are complicated in general, a simple and special solution is given by 
\begin{equation}
P^{abc}(w,z)=f^{abc}P(w,z),
\end{equation}
\noindent
where $f^{abc}$ is a totally symmetric constant and $P(w,z)=\hat{\Delta}(1/(wz))\hat{M}(w,z)$
 is a solution  of 
single component CLR  (\ref{hat-CLR}). This means 
\begin{equation}
\hat{M}^{abc}(w,z)=\Gamma^{ad}f^{dbc}\hat{M}(w,z).
\end{equation}

We comment here that when the above  indices $a,b,c$ are somehow interpreted  as spinorial indices (like dimensional reduction to 1-dim.), 
the Wilson and overlap-type\cite{overlap-1,overlap-2} fermions appear in analogous way with higher dimension.
The Wilson fermion case is realized by a doubler-suppressing term $\bar{\psi} H \psi$ where
\begin{equation}
H_{mn}^{ab}=\delta^{ab}ir\frac{2-\delta_{m,n-1}-\delta_{m,n+1}}{2},
\qquad\hat{H}(w)^{ab}=r\frac{2-w-w^{-1}}{2}\delta^{ab} .
\end{equation}
\noindent
The Wilson-Dirac operator is written as
\begin{equation}
i\hat{D}_{{\rm WD}}(w) = \frac{w-w^{-1}}{2}\Gamma + r\frac{2-w-w^{-1}}{2} ,
\end{equation}
\noindent
and  the bosonic inverse propagator with the doubler-suppressing term is
\begin{equation}
\hat{D}_B(w) =  \Big(\frac{w-w^{-1}}{2}\Big)^2
-r^2\Big(\frac{2-w-w^{-1}}{2}\Big)^2 
=-\gamma_5\hat{D}_{{\rm WD}}(w)\gamma_5\hat{D}_{{\rm WD}}(w),
\end{equation}
where $\gamma_5$ is a matrix satisfying $\{ \gamma_5, \Gamma \} =0$ and $\gamma_5^2=1$. 

For the overlap-type fermion, 
we take 
\begin{equation}
\hat{\Delta}(w)=\frac{w-w^{-1}}{2}K_{\rm OD}(w),
\end{equation}
\noindent
\begin{equation}
\hat{M}^{abc}(w,z) =\frac{f^{abc}}{6 K_{\rm OD}(w)}\Big(2wz+wz^{-1}+zw^{-1}+2(wz)^{-1}\Big),
\end{equation}
\noindent
where 
\begin{equation}
K_{\rm OD}(w)=\frac{M+1}{\sqrt{(M+w)(M+w^{-1})}}.
\end{equation}
\noindent
The overlap-Dirac operator is written as
\begin{equation}
 i\hat{D}_{\rm OD}   = \Gamma \hat\Delta(w) + \hat{H}(w).
\end{equation}
\noindent
The Ginsparg-Wilson relation\cite{Ginsparg-Wilson,Neuberger} is written as 
\begin{equation}
\hat{D}_{\rm OD}\gamma_5+ \gamma_5 \hat{D}_{\rm OD}=      \hat{D}_{\rm OD}   
  \gamma_5 \hat{D}_{\rm OD} .
\label{GWR}
\end{equation}
\noindent
This relation leads us to 
\begin{equation}
 \hat\Delta(w) \hat\Delta(w^{-1}) + \hat{H}^2(w) = 2i  \hat{H}(w)  .
\end{equation}
\noindent
Then,  the bosonic inverse propagator with the doubler-suppressing term 
is expressed as 
\begin{equation}
\hat{D}_B(w) = -\Big(  \frac{(1-w^2)K_{\rm OD}(w)}{2w}\Big) + \hat{H}^2(w) =2i \hat{H}(w) .
\end{equation}
\noindent
It is a remarkable fact  that the inverse propagator of a  bosonic variable is  determind by 
the fermion doubler's mass  term  due to the supersymmetry. 

\section{Higher-dimensional case}    
In the previous section we consider a multi-component extension which will give a starting point for higher dimensional case. In this section, we consider a contrary case, namely two dimensional extension of the CLR without spinor index. Although it is not a real higher dimensional extension, we can readily see the difficulty which may exist in the real case. Real higher dimensional extension will be discussed elsewhere.

Let us denote a two dimensional lattice sites by $\bm{m} \equiv (m_1,m_2)$ and complex coordinates for two-variables $w$-expression by $\bm{w}\equiv(w_1,w_2)$.
Then we use the following simplified notations:
\begin{equation}
\bm{w} \bm{z}\equiv(w_1z_1, w_2z_2)\quad{\rm for}\quad
\bm{w}\equiv(w_1,w_2)\quad{\rm and}\quad\bm{z}=(z_1,z_2),
\end{equation}
also we write
\begin{equation}
{\bf 1}\equiv(1,1)\quad{\rm and}\quad\bm{w}^{-1}\equiv(w_1^{-1},w_2^{-1}). 
\end{equation}
\noindent

Now we consider the following CLR for any bosonic fields 
\begin{equation}
(\bm{\Delta}\Phi,\{ \Psi,   \Xi  \}) + (\bm{\Delta}\Psi,\{ \Xi,   \Phi  \})
+ (\bm{\Delta}\Xi,\{ \Phi,   \Psi  \}) =0,
\label{2-dim-CLR-1}
\end{equation}
\noindent
where the two-dimensional difference operator and the symmetric product are defined by
\begin{equation}
\bm{\Delta} \equiv (\Delta_1,\Delta_2) ,\qquad
\{ \Psi,   \Xi   \}_{\bm{k}} \equiv \sum_{\bm{m}\bm{n}}M_{\bm{k}\bm{m}\bm{n}}\Psi_{\bm{m}} 
 \Xi_{\bm{n}} .
\end{equation}
\noindent
Here $\Delta_1$ and $\Delta_2$ operate the first and second site index $(m_1,m_2)$ of the field respectively.
By imposing  the  translational invariance, 
\begin{equation}
\bm{\Delta}_{\bm{m}\bm{n}} = \bm{\Delta}(\bm{m}-\bm{n}),\qquad
M_{\bm{k}\bm{m}\bm{n}}=M(\bm{k}-\bm{m},\bm{k}-\bm{n}) ,
\end{equation}
\noindent
two  holomorphic functions can be defined as two-variables $w$-expression
\begin{equation}
\hat{\bm{\Delta}}(\bm{w}) \equiv \sum_{\bm{n}}\bm{w}^{\bm{n}} \bm{\Delta}(\bm{n}),
\quad\hat{\bm{\Delta}}(\bm{1})=0, 
\end{equation}
\begin{equation}
\hat{M} (\bm{w},\bm{z}) \equiv \sum_{\bm{m},\bm{n}} \bm{w}^{\bm{m}}\bm{z}^{\bm{n}}
M(\bm{m},\bm{n}) = \hat{M} (\bm{z},\bm{w}) , 
\end{equation}
\noindent
where simplified notation $\bm{w}^{\bm{m}} \equiv w_1^{m_1}w_2^{m_2}$ is used. 
Then, the $w-$expression of the CLR  can be  obtained  as
\begin{equation}
\hat{\bm{\Delta}}(\bm{w}^{-1}\bm{z}^{-1})\hat{M}(\bm{w},\bm{z})+
\hat{\bm{\Delta}}(\bm{w})\hat{M}(\bm{z},\bm{w}^{-1}\bm{z}^{-1})+
\hat{\bm{\Delta}}(\bm{z})\hat{M}(\bm{w}^{-1}\bm{z}^{-1},\bm{w})=0  .
\label{2-dim-CLR-2}
\end{equation}
\noindent
Setting $\bm{z}=\bm{1}$,  we obtain
\begin{equation}
\hat{\bm{\Delta}}(\bm{w}^{-1}) =
 - \frac{\hat{M}(\bm{1},\bm{w}^{-1})}{\hat{M}(\bm{1},\bm{w})}
 \hat{\bm{\Delta}}(\bm{w}) ,
\end{equation}
\noindent
and 
\begin{equation}
\hat{\bm{\Delta}}(\bm{w}\bm{z})= \frac{\hat{M}(\bm{1},\bm{wz})}{\hat{M}(\bm{w},\bm{z})
\hat{M}(\bm{1},(\bm{wz})^{-1})}
\Big( \hat{M}(\bm{z},(\bm{wz})^{-1})\hat{\bm{\Delta}}(\bm{w})  
+ \hat{M}(\bm{w},(\bm{wz})^{-1})\hat{\bm{\Delta}}(\bm{z}) \Big) .
\label{2-difference-1}
\end{equation}
\noindent
From eq.(\ref{2-difference-1}), we have
\begin{equation}
\hat{\bm{\Delta}}(\bm{w}^2)= 2\frac{\hat{M}(\bm{1},\bm{w}^2)}{\hat{M}(\bm{w},\bm{w})
\hat{M}(\bm{1},(\bm{w})^{-2})}
\hat{M}(\bm{w},(\bm{w})^{-2})\hat{\bm{\Delta}}(\bm{w})  .
\label{2-difference-2}
\end{equation}
\noindent
By the use of the Weierstrass preparation theorem\cite{hormander} on multi-variable analytic functions,
the above difference operator can be parametrized as
\begin{equation}
\hat{\bm{\Delta}}(\bm{w}) = \bm{K}(\bm{w})\Big( w_1-1 + (w_2-1)u(w_2) \Big)  , 
\label{2-diff-param}
\end{equation}
\noindent
where $\bm{K}$ and $u$ are nonzero and holomorphic near $\bm{w} \sim \bm{1}$. 

In the following, we shall prove the CLR (\ref{2-dim-CLR-2}) has only a solution 
$\hat{\bm{\Delta}}(\bm{w}) \propto w_1w_2 -1$ with  
the local difference operator $\bm{\Delta}$ and the local product coefficient $M$. 
Since also in this two-dimensional case, the holomorphism is equivalent to the locality, we discuss the relation (\ref{2-dim-CLR-2}) based on 
a theory of  multi-variable holomorphic functions.

Using eqs. (\ref{2-difference-1}),(\ref{2-difference-2}) repeatedly, 
we obtain the following recursion formula, 
\begin{equation}
\hat{\bm{\Delta}}(\bm{w}^n)=n  F(\bm{w},n)\hat{\bm{\Delta}}(\bm{w}) ,
\label{recursion}
\end{equation}
\noindent
where $F$ is a holomorphic function of $\bm{w}$ and 
becomes unity in the naive continuum limit, $\bm{w} \rightarrow \bm{1}$. 
From the parametrization $(\ref{2-diff-param})$, the recursion formula is rewritten by 
\begin{eqnarray}
 w_1^n-1 + (w_2^n-1)u(w_2^n)&=& 
 \frac{nF(\bm{w},n) K(\bm{w})}{ K(\bm{w}^n)}\Big( w_1-1 + (w_2-1)u(w_2) \Big) \nonumber \\
 &=& \prod_{k=1}^n \Big( w_1-w^{(k)}(w_2^n)  \Big)  ,
\label{factor-1}
\end{eqnarray}
\noindent
where 
\begin{equation}
w^{(k)}(w_2^n) \equiv \omega_n^k\Big(   1-  (w_2^n -1)u(w_2^n)
 \Big)^{\frac{1}{n}},~~~~\omega_n=e^{\frac{2i\pi}{n}}  .
\end{equation}

\noindent
From the uniquness of the factorization near the naive continuum limit $\bm{1}$, 
\begin{equation}
w_1-w^{(n)}(w_2^n) =  w_1-1 + (w_2-1)u(w_2)
\label{factor-2}
\end{equation}
\noindent 
because any considered function is holomorphic around the limit. 
The equation (\ref{factor-2}) can be exactly solved as 
\begin{equation}
u(w_2)= \frac{1-w_2^{-u(1)}}{w_2-1} .
\end{equation}
\noindent
Imposing the single-valued condition on $u(w_2)$, $u(1)$ must be an integer, $\ell$. 
The resulting expression is 
\begin{equation}
\hat{\bm{\Delta}}(\bm{w}) = \bm{K}(\bm{w})(w_1-\frac{1}{w_2^{\ell}}) 
= \frac{\bm{K}(\bm{w})}{w_2^{\ell}}(w_1w_2^{\ell}-1    ) .
\end{equation}
\noindent
Instead of $w_1$, we can argue the same thing about $w_2$ of $\hat{\bm{\Delta}}(\bm{w})$. 
Since  both cases hold, $\ell$ must be unity and we can obtain a general form of 
two-dimensional difference operator,
\begin{equation}
\hat{\bm{\Delta}}(\bm{w}) = \bm{K}(\bm{w})(w_1-\frac{1}{w_2}) 
= \frac{\bm{K}(\bm{w})}{w_2}(w_1w_2-1    ) .
\label{2-diff-op}
\end{equation}
\noindent
This general form (\ref{2-diff-op}) is a trivial extension of 
one-dimensional difference operators and the general solution 
 of CLR can be found easily 
by substitution: $w\rightarrow w_1w_2, ~z\rightarrow z_1z_2$ in one-dimensional case. 
This solution is, however, unsuitable for euclidean field theory because, with this,
the kinetic term  is  always written as a form of $w_1w_2$ or, in the momentum representation, $p_1+p_2$.

\vspace{1 cm}

\section{Conclusion and discussion}

When we naively tried to realize the full supersymmetry on lattice, we were required to realize a Leibniz rule (LR) on lattice. The no-go theorem, however, prevented us from proceeding further. In this sense, the theorem is indeed an origin of the difficulty in lattice supersymmetry~\cite{KSS,bergner-1,bergner-2}. On the contrary, realizing exact half-supersymmetry leads us a new criterion, {\it i.e.}~the cyclic Leibniz rule (CLR). The important thing is that the CLR actually has solutions which are realized by the set of local difference operator and local symmetric field products. Unlike the LR, the CLR do not contradict with the locality and the translational invariance on lattice. Yet in the naive continuum limit, CLR cannot be distinguished with LR. Because of these features we consider the CLR may be an important key ingredient for further constructing supersymmetric lattice models beyond the quantum mechanics model discussed in this paper.

Using a difference operator and field products which satisfy the CLR, we can easily construct exact half-supersymmetric action with superfield formalism. We can also find the Nicolai map without ``surface term''. Since each term of our superspace action is independently invariant under supersymmetry, we can introduce deformation parameters with no difficulty. This enable us to utilize a localization technique to compute, for  instance, the Witten index of our supersymmetric quantum mechanics models, which reproduces the known results in the continuum.

In order to construct supersymmetric field theoretical model, we have to adapt spinor index for higher dimensional CLR. In this paper, as a first step, we consider two-dimensional CLR without spinor index, in which we find a difficulty that the difference operator has a particular form unsuitable for the kinetic term. We hope to report a further analysis which properly includes spinor index elsewhere.

\vspace{5mm}
\vspace{10mm}
%
%
%
%
%
%
\acknowledgments
This work is supported in part by the Grant-in-Aid for Scientific 
Research (No.20340048 (M.K.), No.22540281(M.S.) and No.20540274 (H.S.))
by the Japanese Ministry of Education, Science, Sports and Culture.


\vspace{10mm}
%
%
%
\appendix
\section{General solution of the CLR and the properties of 
 smeared symmetric product}\label{appendix_solution}
%
%
%

A general  difference operator is written as a function of the difference of site index
\begin{equation}
\Delta_{mn} = \Delta(m-n), 
\end{equation}
\noindent
by the translational invariance. 
The momentum representation ($w$-expression) is  written as a holomorphic function, 
\begin{equation}
\hat{\Delta}(w) \equiv \sum_{n=-\infty}^{\infty} w^n \Delta(n),
\end{equation}
\noindent
where $w=\exp(ipa)$ in terms of the momentum $p$ dual to the coordinate $na$. This is holomorphic in the domain ${\cal D}$ which is an annulus including  a unit circle 
around the origin. 
From (\ref{diff}), the constraint
\begin{equation}
\hat{\Delta}(1)=0 
\end{equation}
is imposed on the difference operator. 
For the  product coefficient between lattice fields, 
we can also write in the translational invariant  form 
\begin{equation}
M_{\ell mn} = M(\ell-m,\ell-n) .
\label{s-product}
\end{equation}
\noindent
The momentum representation is  written as a two-variables holomorphic function, 
\begin{equation}
\hat{M}(w,z) \equiv \sum_{m,n=-\infty}^{\infty} w^mz^n M(m,n) = \hat{M}(z,w),
\end{equation}
\noindent
where $w=exp(ipa), ~z=exp(iqa)$ and the holomorphic domain is  ${\cal D}\otimes{\cal D}$.
Then, our cyclic  Leibniz rule (\ref{symmetric Leibniz}) can be expressed as 
\begin{equation}
\hat{M}(w,z)\hat{\Delta}(\frac{1}{wz})+\hat{M}(z,\frac{1}{wz})\hat{\Delta}(w)+\hat{M}(\frac{1}{wz},w)\hat{\Delta}(z)=0 .
\end{equation}
\noindent
The general solution for this equation is 
\begin{eqnarray*}
 \hat{M}(w,z)\hat{\Delta}(\frac{1}{wz})&=&  f(w,z)+f(z,1)+ f(wz,\frac{1}{wz})
 +f(z,w)+f(w,1)+   f(wz,\frac{1}{wz})\\
 &-&f(z,\frac{1}{wz}) -f(\frac{1}{wz},1)-f(\frac{1}{w},w) 
 -f(w,\frac{1}{wz}) -f(\frac{1}{wz},1)-f(\frac{1}{z},z),
 \label{general-sol}
\end{eqnarray*}
\noindent
where $f(w,z)$ is an arbitrary two-variable holomorphic function. 
By taking $f(w,z)=F(w)+F(z)$, the special solution is found 
\begin{equation}
 \hat{M}(w,z)\hat{\Delta}(\frac{1}{wz})=F(w)-F(\frac{1}{w})
 +F(z)-F(\frac{1}{z})-2F(\frac{1}{wz})+2F(wz) .
\end{equation}
\noindent
If we choose $F(w)=\frac{w^2}{12}$, then 
\begin{equation}
 \hat{M}(w,z)\hat{\Delta}(\frac{1}{wz})=\frac{2wz+z^{-1}w+wz^{-1}+2(wz)^{-1}}{6}\frac{wz-(wz)^{-1}}{2} .
\end{equation}
\noindent
This corresponds to the momentum representation of (\ref{eg-1}). 

Note that we cannot impose an associative law on our symmetric product.
Namely,   
 the product (\ref{s-product}) cannot satisfy the associative law 
\begin{equation}
\{\phi,\{\psi,\chi \}  \}=  \{ \{\phi, \psi \} , \chi \}. 
\label{associative-law}
\end{equation}
\noindent
If we imposed (\ref{associative-law}), then 
\begin{equation}
\sum_{\ell} M_{\ell mn} M_{j\ell k}  = \sum_{\ell} M_{jm\ell}M_{\ell nk}
\label{associative-law2}
\end{equation}
\noindent
or in $w$-expression, 
\begin{equation}
\hat{M}(w,z)\hat{M}(wz,x)=\hat{M}(z,x)\hat{M}(xz,w)
\label{associative-law3}
\end{equation}
\noindent
would hold which implied $\hat{M}(w,z)=1$ after local field redefinition\cite{KSS},
while the  constant   $\hat{M}$ was excluded from our general solution(\ref{general-sol}) .  

Another property of the symmetric product (\ref{s-product}) may come out
for a case where there exists a unit element $I$ which satisfies
\begin{equation}
\{\phi,I\}=\{I,\phi\}=\phi.
\end{equation}
Since $I_n$ can be taken as a unit constant by a local field redefinition,
combining with the translational invariance, the above implies 
\begin{equation}
\hat{M}(w,1)=\hat{M}(1,w)=1
\label{unit}
\end{equation}
\noindent
for any $w$.  Using the solution (\ref{general-sol}),  
we can find an example satisfying (\ref{unit}), 
\begin{equation}
\hat{M}(w,z)=1+\frac{1}{12wz}(w-1)(z-1)(w+z-2-2wz). 
\end{equation}


\section{Multi-body products and the CLR}

In this appendix we describe how multi-body smeared symmetric products satisfying CLR can be constructed from 2-body ones.

\subsection*{B.1 Construction}
$N$-body smeared symmetric product $\{\phi^{(1)},\phi^{(2)},\cdots,\phi^{(N)}\}$ is defined as
\begin{equation}
\{\phi^{(1)},\phi^{(2)},\cdots,\phi^{(N)}\}_n
=\sum_{n_1,\cdots,n_N} M_{n;n_1\cdots n_N}
\phi^{(1)}_{n_1}\phi^{(2)}_{n_2}\cdots \phi^{(N)}_{n_N},
\end{equation}
with the coefficient $M_{n;n_1\cdots n_N}$ whose last $N$ indices are totally symmetric. The CLR relation for this can be expressed as
\begin{equation}
\left(\Delta\phi,\{\phi,\cdots,\phi\}\right)=0
\end{equation}
for a single field $\phi$, or
\begin{equation}
\sum_{\{c(i) \mbox{ : cyclic perm.}\}}
\left(\Delta\phi^{c(1)},\{\phi^{c(2)},\cdots,\phi^{c(N+1)}\}\right)=0
\end{equation}
for $N+1$ fields $\phi^i$ ($i=1,2,\cdots,N+1$), where the sum is taken over all possible cyclic permutations $\{c(i)\, |\, i=1,\cdots,N+1\}$. 

$N$-body symmetric product can be constructed from 2-body product by the following steps:
\begin{enumerate}
\item Write down a tree diagram (without loops) with $N+1$ external lines consists only of trivalent vertices. (See Fig.~\ref{clr-fig1} for an $N=4$ example.) 

\begin{figure}[htbp] 
\begin{center}
\includegraphics[scale=.4]{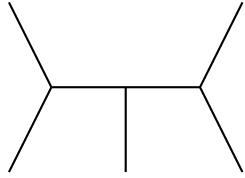}
\caption{Example of a tree diagram from which one can construct 4-body product}
\label{clr-fig1}
\end{center}
\end{figure}         

\item Choose one of the external lines and assign it ``0". Then number the rest of external lines from 1 to $N$ in arbitrary order. (See Fig.~\ref{clr-fig2} for counter-clockwise order examples)

\item For each vertex mark a dot on a line, among three lines attached to the vertex, which is connected to the portion containing ``0"-line. (See Fig.~\ref{clr-fig2}) Now this diagram expresses one of 4-body products composed of three 2-body products. Each vertex corresponds to 2-body product coefficient $M_{nml}$ and a dot distinguishes first index. Indices of internal lines (lines connecting vertices) are summed. Thus, for example, Fig.~\ref{clr-fig2}(a) corresponds to $\sum_{k,l}M_{nn_1k}M_{kn_2l}M_{ln_3n_4}$ which gives a product of fields
$\{\phi^{(1)},\{\phi^{(2)},\{\phi^{(3)},\phi^{(4)}\}\}\}$.

\begin{figure}[htbp] 
\begin{center}
\includegraphics[scale=.3]{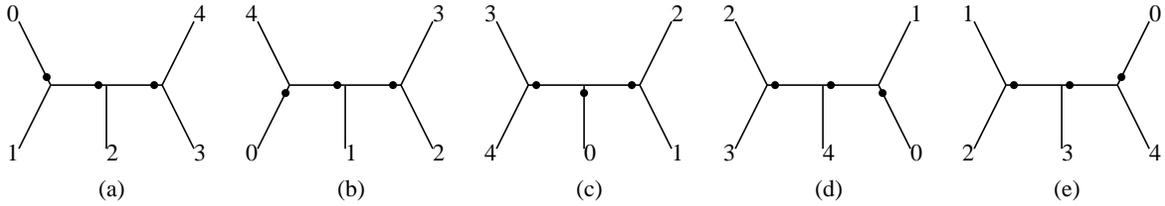}
\caption{Constructing 4-body product made of 2-body products}
\label{clr-fig2}
\end{center}
\end{figure}         

\item For all possible choices of ``0"-line and any permutatios of ``1" through ``$N$", repeat the steps 2 and 3. Then sum up all combined products corresponding to these diagrams with normalization factor $1/(N+1)!$. 
For $N=4$ case, all possible choices of ``0"-line obtained from Fig.~\ref{clr-fig1} are shown in Fig.~\ref{clr-fig2}, and the final expression is
\begin{eqnarray}
\hspace{-8mm}
M_{n;n_1n_2n_3n_4}&=&\frac{1}{5!}\sum_{\mbox{perm.}}\sum_{k,l}\left(
M_{nn_{p(1)}k}M_{kn_{p(2)}l}M_{ln_{p(3)}n_{p(4)}}
+M_{kn_{p(1)}l}M_{ln_{p(2)}n_{p(3)}}M_{nkn_{p(4)}}\right.\nonumber\\
&&+\left.M_{nkl}M_{kn_{p(1)}n_{p(2)}}M_{ln_{p(3)}n_{p(4)}}
+M_{nn_{p(1)}k}M_{ln_{p(2)}n_{p(3)}}M_{kln_{p(4)}}\right.\nonumber\\[5pt]
&&+\left.M_{ln_{p(1)}n_{p(2)}}M_{kln_{p(3)}}M_{nkn_{p(4)}}\right).
\end{eqnarray}
This gives 4-body field product as
\begin{eqnarray}
\hspace{-8mm}\{\phi^{(1)},\phi^{(2)},\phi^{(3)},\phi^{(4)}\}
&=&\frac{1}{5!}\sum_{\mbox{perm.}}
\left(\{\phi^{(p(1))},\{\phi^{(p(2))},\{\phi^{(p(3))},\phi^{(p(4))}\}\}\}\right.\nonumber\\
&&\hspace{-30mm}+\{\{\phi^{(p(1))},\{\phi^{(p(2))},\phi^{(p(3))}\}\},\phi^{(p(4))}\}
+\left.\{\{\phi^{(p(1))},\phi^{(p(2))}\},\{\phi^{(p(3))},\phi^{(p(4))}\}\}\right.\nonumber\\[7pt]
&&\hspace{-30mm}+\{\phi^{(p(1))},\{\{\phi^{(p(2))},\phi^{(p(3))}\},\phi^{(p(4))}\}\}
+\left.\{\{\{\phi^{(p(1))},\phi^{(p(2))}\},\phi^{(p(3))}\},\phi^{(p(4))}\}\right).\label{4-body_product}\hspace{5mm}
\end{eqnarray}
\end{enumerate}
Thus we obtain one of possible definitions of multi-body symmetric product for each tree diagram we start from.
Although tree diagram for $N=3$ or 4 case is unique, there are a finite number of independent tree diagrams for higher $N$. Each of these diagrams gives independent product on a lattice, while they may have the same continuum limit.

Note that a class of multi-body symmetric products can be alternatively obtained by a simple equation
\begin{equation}
\{\phi,\phi,\cdots,\phi\}\equiv
\frac{1}{N+1}\sum_{k=0}^N\{P_{N_k}(\phi),P_k(\phi)\}.
\end{equation}
Here the field $P_n(\phi)$ is defined in a recursive way using 2-body product
\begin{equation}
P_{n}(\phi)=\{P_{n-1}(\phi),\phi\},\qquad P_1(\phi)=\phi,\qquad P_0(\phi)=I
\end{equation}
where $I$ is an identity field with respect to the 2-body product ($\phi=\{I,\phi\}$).

\subsection*{B.2 Proof of multi-body CLR}
Here we show that the above constructed products satisfy the multi-body CLR.
A graphical representation is convenient for this purpose. As in the previous subsection, coefficient of 2-body product $M_{nml}$ is expressed by the trivalent vertex with a dot on the first index. Difference operator $\Delta_{nm}$ is now expressed by a line segment with an arrow:
\begin{equation}
M_{nml}\qquad
\begin{picture}(30,30)(0,-3)
\put(0,0){\line(1,0){23}}
\put(23,0){\line(1,2){10}}
\put(23,0){\line(1,-2){10}}
\put(19,0){\circle*{3}}
\put(-8,-3){\mbox{$n$}}
\put(35,20){\mbox{$l$}}
\put(35,-23){\mbox{$m$}}
\end{picture}
\qquad\qquad\qquad\qquad
\Delta_{nm}\qquad
\begin{picture}(30,30)(0,-3)
\put(0,0){\line(1,0){28}}
\put(10,0){\line(2,1){10}}
\put(10,0){\line(2,-1){10}}
\put(-8,-3){\mbox{$n$}}
\put(30,-3){\mbox{$m$}}
\end{picture}
\vspace{6mm}
\end{equation}

Anti-symmetry of $\Delta$ ($=-\Delta^T$) is represented as a graphical relation:
\begin{equation}
\begin{picture}(30,30)(50,-10)
\put(0,0){\line(1,0){28}}
\put(10,0){\line(2,1){10}}
\put(10,0){\line(2,-1){10}}
\put(-8,-3){\mbox{$n$}}
\put(30,-3){\mbox{$m$}}
\put(50,-3){\mbox{$=\  -$}}
\put(90,0){\line(1,0){28}}
\put(108,0){\line(-2,1){10}}
\put(108,0){\line(-2,-1){10}}
\put(82,-3){\mbox{$n$}}
\put(120,-3){\mbox{$m$}}
\end{picture}
\vspace{3mm}
\end{equation}
Then CLR for 2-body product
$\sum_{\mbox{cyclic}}\left(\Delta\phi^{c(1)},\{\phi^{c(2)},\phi^{c(3)}\}\right)=0$ is graphically represented by Fig.~\ref{clr-fig3}.
\begin{figure}[htbp] 
\begin{center}
\includegraphics[scale=.4]{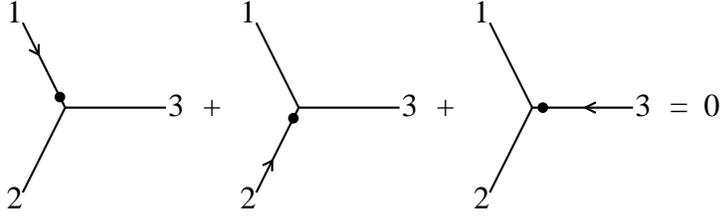}
\caption{CLR for 2-body products}
\label{clr-fig3}
\end{center}
\end{figure}         

Let us begin by looking at $N=4$ example. What we want to show is the relation
\begin{equation}
\sum_{c\,:\,\mbox{cyclic perms.}}
\left(\Delta\phi^{c(1)},
\{\phi^{c(2)},\phi^{c(3)},\phi^{c(4)},\phi^{c(5)}\}\right) = 0.
\label{4-body_CLR}
\end{equation}
Substituting the definition of 4-body product (\ref{4-body_product}) into the left-hand side of (\ref{4-body_CLR}), we have 600 terms in total ($5!$ kinds of terms times 5 cyclic permutations). These can be rearranged into 5! groups, one of which made of five graphs shown in Fig.~\ref{clr-fig4} and the others are its permutations with respect to external lines. Therefor it is enough to argue the cancellations of graphs in one group. Actually contributions of five graphs in Fig.~\ref{clr-fig4} add up to zero by the CLR for 2-body product (Fig.~\ref{clr-fig3}) and the anti-symmetry of $\Delta$.
\begin{figure}[htbp] 
\begin{center}
\includegraphics[scale=.3]{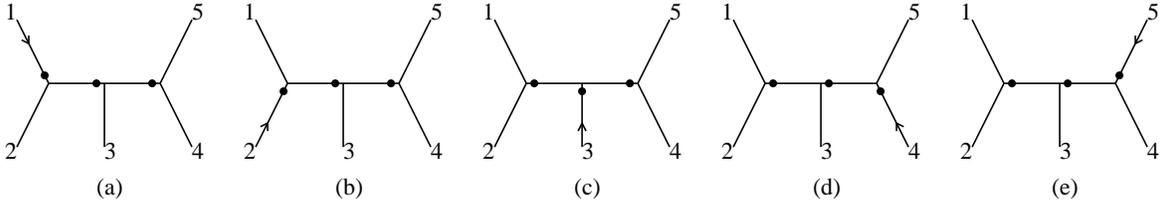}
\caption{Vanishing set of graphs in CLR for 4-body products}
\label{clr-fig4}
\end{center}
\end{figure}         
\begin{figure}[htbp] 
\begin{center}
\includegraphics[scale=.3]{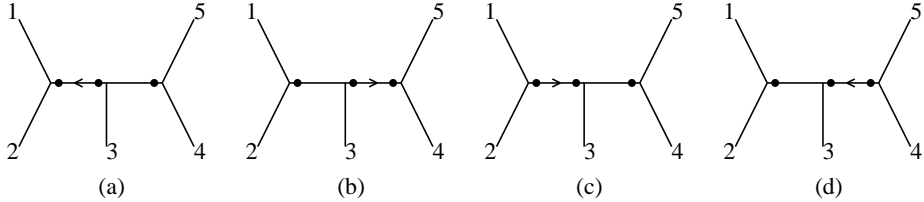}
\caption{Canceling graphs: (a)+(c)=0 and (b)+(d)=0 due to $\Delta+\Delta^T=0$}
\label{clr-fig5}
\end{center}
\end{figure}         
For example, first two graphs in Fig.~\ref{clr-fig4} add up to become the graph Fig.~\ref{clr-fig5}(a) with minus sign by the 2-body CLR, while last two graphs in Fig.~\ref{clr-fig4} do become the graph Fig.~\ref{clr-fig5}(b) again with minus sign. Remaining Fig.~\ref{clr-fig4}(c), however, can be rewritten into a sum of the graphs Fig.~\ref{clr-fig5}(c) and (d) with minus signs. Therefor they cancel out altogether due to anti-symmetry of $\Delta$.

We can generalize the above arguments to arbitrary $N$ case by induction.
Suppose we already have the $M$-body CLR for $M$ less than $N$. Consider a tree graph with $N+1=K+L$ external lines (See left hand side of Fig.~\ref{clr-fig6}). We can find an internal line which separates the graph into two parts each of which has $K$ and $L$ external lines respectively (See right hand side of Fig.~\ref{clr-fig6}), where $1<K<N$ and $1<L<N$.
\begin{figure}[htbp] 
\begin{center}
\includegraphics[scale=.4]{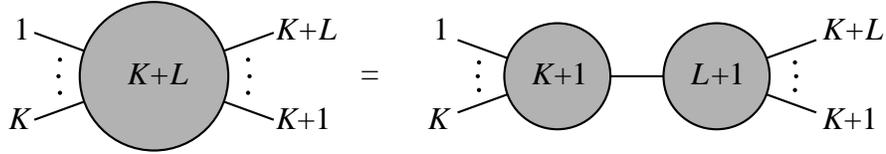}
\caption{Factorization into products with smaller numbers}
\label{clr-fig6}
\end{center}
\end{figure}         
To examine the CLR, we attach $\Delta$ at one of external lines and sum up each contributions. Then the contributions of the graphs attached $\Delta$ at an external line with number 1 though $K$ are turned out to be that of Fig.~\ref{clr-fig7}(a) with minus sign by the $K$-body CLR, while those with number $K+1$ through $K+L$ be that of Fig.~\ref{clr-fig7}(b) with minus sign by the $L$-body CLR. As was the case of $N=4$, Fig.~\ref{clr-fig7}(a) and (b) cancel each other by the anti-symmetry of $\Delta$. Thus $N$-body CLR is satisfied.
\begin{figure}[htbp] 
\begin{center}
\includegraphics[scale=.4]{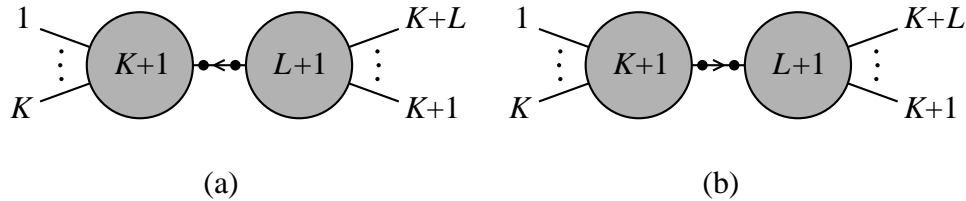}
\caption{Canceling graphs: (a)+(b)=0 due to $\Delta+\Delta^T=0$}
\label{clr-fig7}
\end{center}
\end{figure}         

\vspace{10mm}
%
%
%
%
%
%

%
%
%
%
%
%
\end{document}